\documentclass[12pt]{article}
\usepackage{amsmath,amssymb,amsthm}
\newtheorem{Thm}{Theorem}
\usepackage{epsfig}

\newcommand{\bfm}[1]{\mbox{\boldmath ${#1}$}}
\newcommand{\nonum}{\nonumber \\}
\newcommand{\beqa}{\begin{eqnarray}}
\newcommand{\eeqa}[1]{\label{#1}\end{eqnarray}}
\newcommand{\beq}{\begin{equation}}
\newcommand{\eeq}[1]{\label{#1}\end{equation}}
\newcommand\eq[1] {(\ref{#1})} 
\newcommand{\Grad}{\nabla}

\newcommand{\Tr}{\mathop{\rm Tr}\nolimits}

\newcommand{\Md}{\partial}

\newcommand{\Ga}{\alpha}
\newcommand{\Gb}{\beta}

\newcommand{\Ge}{\epsilon}

\newcommand{\Gg}{\gamma}

\newcommand{\Gl}{\lambda}

\newcommand{\Gt}{\theta}

\newcommand{\Gs}{\sigma}

\newcommand{\GO}{\Omega}

\newcommand{\BGe}{\bfm\epsilon}

\newcommand{\BGs}{\bfm\sigma}

\newcommand{\CC}{{\cal C}}

\newcommand{\CF}{{\cal F}}

\newcommand{\bpm}{\begin{pmatrix}}
\newcommand{\epm}{\end{pmatrix}}
\newcommand\fig[1] {{\rm Figure}~\ref{fig:#1}}

\newcommand\labfig[1] {\label{fig:#1}}

\usepackage{url}
\usepackage{hyperref}

\def\Bf{{\bf f}}

\def\Bs{{\bf s}}
\def\Bt{{\bf t}}
\def\Bu{{\bf u}}

\def\Bx{{\bf x}}
\def\By{{\bf y}}

\def\BF{{\bf F}}

\def\BR{{\bf R}}

\def\BT{{\bf T}}

\def\BX{{\bf X}}

\usepackage{graphics}
\usepackage{amssymb}

\begin{document}
\title{The set of forces that ideal trusses, or wire webs, under tension can support}
\author{Graeme W. Milton}
\date{\small{Department of Mathematics, University of Utah, Salt Lake City, UT 84112, USA
\\Email: milton@math.utah.edu}}
\maketitle
\vspace{2ex}
\begin{abstract}
The problem of determining those multiplets of forces, or sets of force multiplets, acting at a set of points, such that
there exists a truss structure, or wire web, that can support these force multiplets with all the elements of the truss
or wire web being under tension, is considered.
The two-dimensional problem where the points are at the vertices of a convex polygon is essentially
solved: each multiplet of forces must be such that the net anticlockwise torque around
any vertex of the forces summed over any number of consecutive points clockwise past the vertex must be
non-negative; and one can find a truss structure that supports under tension, and only supports, those
force multiplets in a convex polyhedron of force multiplets that is generated by a finite number of
force multiplets each satisfying the torque condition. Progress is also made on the problem
where only a subset of the points are at the vertices of a convex polygon, and the other points are inside.
In particular, in the case where only one point is inside, an explicit procedure is described for
constructing a suitable truss, if one exists. An alternative
recipe to that provided by  Guevara-Vasquez, Milton, and Onofrei \cite{Vasquez:2011:CCS},
based on earlier work of Camar Eddine and Seppecher \cite{Camar:2003:DCS}, is given for constructing
a truss structure, with elements under either compression or tension, that supports an arbitrary collection of 
balanced forces at the vertices of a convex polygon. Finally some constraints are given on the forces that a three-dimension
truss, or wire web, under tension must satisfy.

\end{abstract}
\vspace{3ex}
\vspace{3ex}
\section{Introduction}
\setcounter{equation}{0}

The analysis of truss structures is fundamental to structural engineering, and a 
good historical account of the subject can be found in the excellent book of Timoshenko \cite{Timoshenko:1983:HSM},
and a more detailed mathematical treatment is given, for example, in the paper of Pellegrino and Calladine \cite{Pellegrino:1986:MAS}.
If one wants to prevent buckling in truss structures it is obviously beneficial to have all
elements under tension. The question is then: given a set of points $\Bx_1,\Bx_2,\ldots,\Bx_n$
and forces $\Bt_1,\Bt_2,\ldots,\Bt_n$ acting at them, can one find a truss structure that supports 
these forces such that all the elements of the truss 
structure are under tension? We will call the forces a force multiplet
$\BF= (\Bt_1,\Bt_2,\ldots,\Bt_{n-1})$ where it is not necessary to keep track
of $\Bt_n$ since by balance of forces 
\beq \Bt_n=-(\Bt_1+\Bt_2+\ldots+\Bt_{n-1}). \eeq{0.0}
Additionally, by balance of torques in the two-dimensional case, we have
\beq \sum_{i=1}^n\Bx_i\cdot[\BR_\perp\Bt_i]=0, \eeq{0.0a}
where 
\beq \BR_\perp=\bpm 0 & 1 \\ -1 & 0 \epm \eeq{1.2}
is the matrix for a $90^\circ$ clockwise rotation. If \eq{0.0} and \eq{0.0a} both hold we will
say that the set of forces is balanced.

We will say the truss structure supports $\BF$ under tension if all the truss members in the truss
structure are under tension. For a fixed set of points $\Bx_1,\Bx_2,\ldots,\Bx_n$,
we can also consider the grander question of characterizing those sets $\CF$ of force multiplets,
such that one can find single truss structure supporting under tension all of them, but none outside $\CF$.

In the two-dimensional case of planar truss structures, where $\Bx_1,\Bx_2,\ldots,\Bx_n$ are at the vertices of a
convex polygon, we obtain an essentially complete answer to both questions. We prove in Section 2 the following theorem:
\begin{Thm}
A set of points $\Bx_1,\Bx_2,\ldots,\Bx_n$ at the vertices of a convex polygon, numbered clockwise, can support balanced
forces $\Bt_1,\Bt_2,\ldots,\Bt_n$ at these vertices, 
with a truss with all its elements under tension, if and only if for all $i$ and $j$,
\beq  \sum_{k=j}^{i-1}(\Bx_k-\Bx_j)\cdot[\BR_\perp\Bt_k]\geq 0,
\eeq{0.1}
and we have assumed $i>j$, if necessary by replacing $i$ by $i+n$ and identifying where necessary $\Bx_k$ and $\Bt_k$ with
$\Bx_{k-n}$ and $\Bt_{k-n}$.
\end{Thm}
The necessary and sufficient condition \eq{0.1} has a physical interpretation: the net anticlockwise torque around
the point $\Bx_j$ of the forces $\Bt_k$ summed over any number of consecutive points clockwise past the point
$\Bx_j$ is non-negative. Similarly, as also implied by \eq{0.1} the net clockwise torque around
the point $\Bx_j$ of the forces $\Bt_k$ summed over any number of consecutive points anticlockwise past the point
$\Bx_j$ is non-negative. If \eq{0.1} is satisfied we provide an explicit recipe for constructing a truss structure
that supports the forces, with all truss elements being under tension.

Let $\CF_0$ be the set of all force multiplets $\BF= (\Bt_1,\Bt_2,\ldots,\Bt_{n-1})$ satisfying \eq{0.1}, where $\Bt_n$ is
given by \eq{0.0}. This set is a convex cone in the sense that if $\BF_1$ and $\BF_2$ are in $\CF_0$ then 
$\Ga_1\BF_1+\Ga_2\BF_2\in\CF_0$ for all $\Ga_1\geq 0$ and $\Ga_2\geq 0$. In section 3 we establish the theorem:
\begin{Thm}
The set of force multiplets $\BF= (\Bt_1,\Bt_2,\ldots,\Bt_{n-1})$ supported by a truss structure under tension
is necessarily a convex cone. Conversely, given any polyhedral convex cone $\CF$ contained in $\CF_0$,
generated by linear multiples with non-negative coefficients of a finite number of force multiplets
$\BF_1,\BF_2,\ldots,\BF_h$ each satisfying \eq{0.1}, then we can find a truss structure that supports under tension, and only supports,
those force multiplets $\BF$ in $\CF$.
\end{Thm}

We also consider, in Section 4, the case where the points $\Bx_1,\Bx_2,\ldots,\Bx_n$ do not all lie at the vertices of a convex polygon. A condition like \eq{0.1}
still holds, but it does not always guarantee that there exists a truss under tension supporting the forces. The case where all
but one of the points lie at the vertices of a convex polygon is explored in depth and in the appendix 
an explicit procedure for constructing a suitable truss structure, if one exists,
is given. The general case, where more than one point lies inside the convex polygon, seems rather knotty and only some suggestions are made in how to construct
an appropriate truss structure.  

Since the elements are all under tension we can consider the equivalent problem of designing a wire network
of inextensible wires that support the desired force multiplet, or set of force multiplets. Also, by changing
the sign of the tensions in each truss element, one sees that the same geometry will support 
$-\BF= (-\Bt_1,-\Bt_2,-\ldots,-\Bt_{n-1})$ with each element being under compression. This too is an important 
problem as some materials, such as concrete, have much greater strength under compression than under tension. 

This work is motivated by an amazing, but not well known, result of Camar Eddine and Seppecher \cite{Camar:2003:DCS} (see their Theorem 5)
who proved by induction that, in three-dimensions, truss structures, with members under either tension or compression,
can support an arbitrary force multiplet $\BF$ and in fact an arbitrary linear subspace $\CF$ of 
force multiplets. (The key to the second result, given the first, is that if one takes a basis
$\BF_1,\BF_2,\ldots,\BF_k$ for $\CF$ and if for each $i=1,2,\ldots, k$ one finds a truss structure that supports
$\BF_i$ and only $\BF_i$, then by superimposing these truss structures, with some minor modifications to avoid collisions
in the structures, one obtains a final structure that supports any force multiplet $\BF\in\CF$, but no multiplet outside $\CF$.)
Their final objective, which they succeeded in doing for three-dimensional linear elasticity (the two-dimensional problem remains open),
was to obtain a complete characterization of the set of all possible local and non-local macroscopic
responses in elastic composites built from isotropic elastic materials with arbitrarily small elastic moduli
and arbitrarily large elastic moduli. This extended the result of Milton and Cherkaev \cite{Milton:1995:WET} that any
positive definite fourth order tensor satisfying the symmetries of elasticity tensors, can in fact
be realized as the effective elasticity tensor of a composite of two isotropic materials, one very
compliant and the other very stiff. 

The result of Camar Eddine and Seppecher was instrumental in the subsequent work of Guevara-Vasquez, Milton, and Onofrei \cite{Vasquez:2011:CCS}
who extended their construction to two-dimensional truss structures (Theorem 2 in \cite{Vasquez:2011:CCS}), and moreover
obtained a complete characterization, within the framework of linear elasticity,
of the dynamic response at the terminal nodes of both  two and three-dimensional mass-spring networks (Theorem 4 in \cite{Vasquez:2011:CCS}). 
In other words, if
in a set of $m$ independent measurements, indexed by $j=1,2,\ldots, m$, one specifies time varying displacements $\Bu_1^{(j)}(t),\Bu_2^{(j)}(t),\ldots,\Bu_n^{(j)}(t)$ at the points
$\Bx_1,\Bx_2,\ldots,\Bx_n$, where $t$ is the time, one can say precisely what resultant force
functions $\Bt_1^{(j)}(t),\Bt_2^{(j)}(t),\ldots,\Bt_n^{(j)}(t)$, $j=1,2,\ldots, m$, (that total $mn$ in number) it is possible to generate at these points, allowing for
an arbitrary number of internal nodes, and an arbitrary mass (possibly zero) at each internal node. In particular,
one can select a desired set of resonant frequencies and one can independently choose any desired response (eigenmode) at each resonant frequency subject only
to natural thermodynamic considerations. The complete characterization and synthesis of the response of mass-spring networks with Rayleigh damping,
where the stiffness matrix, damping matrix and mass matrix are linearly related, was subsequently
obtained by Gondolo and Guevara-Vasquez \cite{Gondolo:2014:CSR}.

Here, in Section 5, we provide an alternative procedure to the inductive one given by Guevara-Vasquez, Milton, and Onofrei \cite{Vasquez:2011:CCS}
(based on the three-dimensional construction of Camar Eddine and Seppecher \cite{Camar:2003:DCS}) for constructing a truss structure 
that supports an arbitrary collection of balanced forces at points $\tilde{\Bx}_1,\tilde{\Bx}_2,\ldots,\tilde{\Bx}_n$ at the vertices of a convex polygon.
The basic idea is to superimpose two truss structures: one with all its elements under compression, and the other with all its 
elements under tension, so that the net forces at the points $\tilde{\Bx}_1,\tilde{\Bx}_2,\ldots,\tilde{\Bx}_n$ are the desired ones.

In Section 6 we briefly visit the difficult problem of three-dimensional wire webs. Again a condition like \eq{0.1} is shown to hold
for every two-dimensional projection of the web, but we make no progress on the problem of actually constructing webs that support
a desired set of forces.

We emphasize that the constraints we provide on a force multiplet supported by a web only rely on the fact that the stress 
has zero divergence and that all elements are under tension. Thus the constraints apply not just to elastic truss structures, but also to
structures with a non-linear elastic response, or possibly to structures under creep or under other plastic deformation, assuming inertial
effects can be neglected. However, the constructions we produce that support a desired set of force multiplets
do not necessarily minimize the elastic compliance energy associated with the truss structure. Thus they do not, for example, necessarily
correspond to a spider web that is in elastic equilibrium, as in that scenario the internal nodes of the web can move to 
minimize the elastic energy. In fact, to obtain the desired constructions, we ignore elastic energy considerations
altogether and treat the struts, or wires, as inextensible. It could be the case that this assumption of inextensiblity can be relaxed,
and that our constructions, with an appropriate choice
of stiffnesses of the struts, can be made elastically stable. However we do not explore this here.  
If one wants truss structures built from a single material that minimize the total compliance energy
then an excellent approach is to use topology optimization methods: see, for example, chapter 4 of Bends{\o}e and Sigmund \cite{Bendsoe:2004:RPM} and references therein. 
If one wants structures, such as formed from concrete reinforced with steel, built from two materials
where one material is strong under tension while the other is strong under compression, then a hybridized truss-continuum topology optimization
may be appropriate \cite{Gaynor:2013:RCF}. Other important considerations may affect the design too: such as making sure the stress in the wires, or struts, 
is not enough to cause damage, or making sure there
are no adverse resonance effects in the frequency range in which they are likely to be excited. We emphasize, too, that the truss or wire networks 
we envisage do not necessarily have any inherent structural rigidity: they need to be attached to the points $\Bx_1,\Bx_2,\ldots,\Bx_n$ where the forces are applied
to give them structural integrity.

\section{Forces at the vertices of a convex polygon}
\setcounter{equation}{0}

Let us consider what set of forces $\Bt_i$, $i=1,2,\ldots,n$, each applied at a
vertex $\Bx_i$ of a convex polygon, can be supported by a web under tension
attached to these points. Here we assume that the vertices are numbered 
clockwise around the polygon, and we identify $\Bt_{n+1}$ and $\Bx_{n+1}$
with $\Bt_1$ and $\Bx_1$. One elementary constraint is rather clear: the vector
$-\Bt_i$ must point inside the convex polygon, or equivalently 
\beq (\Bx_i-\Bx_{i-1})\cdot[\BR_\perp\Bt_i]\geq 0,\quad (\Bx_i-\Bx_{i+1})\cdot[\BR_\perp\Bt_i]\leq 0,
\eeq{1.0}
where $\BR_\perp$ given by \eq{1.2} is the matrix for a $90^\circ$ clockwise rotation.
If this condition does not hold there is no way that
the web wires attached to $\Bx_i$, that are necessarily pulling inside the convex hull, can balance the
force $\Bt_i$. It is also clear that this condition is not sufficient, as the position of the point
$\Bx_{i-1}$ should not matter in the limit in which the force $\Bt_{i-1}$ applied there becomes vanishingly small:
in that limit the correct condition should imply that $(\Bx_i-\Bx_{i-2})\cdot[\BR_\perp\Bt_i]$ is non-negative.

For two-dimensional elasticity it is well known that in the absence of body forces in a simply connected
region $\GO$ the stress field $\BGs(\Bx)$, having zero divergence,
can be represented in terms of the Airy stress function $\phi(\Bx)$:
\beq \BGs(\Bx)=\BR_\perp^T\Grad\Grad\phi(\Bx)\BR_\perp, \eeq{1.1}
in which $\BR_\perp^T=-\BR_\perp$ is the transpose of $\BR_\perp$. This is just a restatement of the fact that the stress tensor
$\BGs(\Bx)$ when expressed in terms of the Airy stress function takes the form
\beq \BGs(\Bx)=\bpm \Gs_{11}(\Bx) &\Gs_{12}(\Bx) \\ \Gs_{21}(\Bx) &\Gs_{22}(\Bx)  \epm 
=\bpm \frac{\Md^2\phi(\Bx)}{\Md^2 x_2^2} &-\frac{\Md^2\phi(\Bx)}{\Md x_1 \Md x_2} \\ -\frac{\Md^2\phi(\Bx)}{\Md x_1 \Md x_2} &
\frac{\Md^2\phi(\Bx)}{\Md^2 x_1^2}\epm,
\eeq{1.1aa}
and if we rotate the matrix on the right by $90^\circ$ we arrive at the double gradient of $\phi(\Bx)$.
Since $\BGs(\Bx)$ is positive semidefinite for all $\Bx$ we
see that $\Grad\Grad\phi(\Bx)$ is positive semidefinite for all $\Bx$
which implies that $\phi(\Bx)$ has non-negative curvature everywhere within the polygon. Thus the Airy stess function is a
convex (or concave) function in any simply-connected two-dimensional region under tension (or compression), that may
have subregions with zero stress \cite{Giaquinta:1988:RMS}.
When $\BGs(\Bx)$ is zero in a region, as it is between the wires in web, then $\Grad\Grad\phi(\Bx)=0$
which implies the Airy stress function is a linear function of $\Bx$  in this region. Thus the Airy stress function associated
with a wire web under tension is a convex polygonal surface with discontinuities of slope along the wires in the web, where
the magnitude of the slope discontinuity can be connected to the tension in the associated wire \cite{Fraternali:2002:LSM,Fraternali:2014:CBF}.

The essential idea behind the following analysis that will prove theorem 1 is shown in \fig{0.5}. Any network under tension that supports
forces at the vertices of a convex polygon as for example in (a), will have an associated Airy stress function that is a convex polyhedron as in (b).
We can replace this convex polyhedron by a simpler convex polyhedron, as in (c), formed from the tangent planes at the boundary of the polygon.
The lines of discontinuity of slope of this simplified Airy stress function then give an equivalent network that supports under tension the same set
of forces as the original network. 

\begin{figure}
	\centering
	\includegraphics[width=0.5\textwidth]{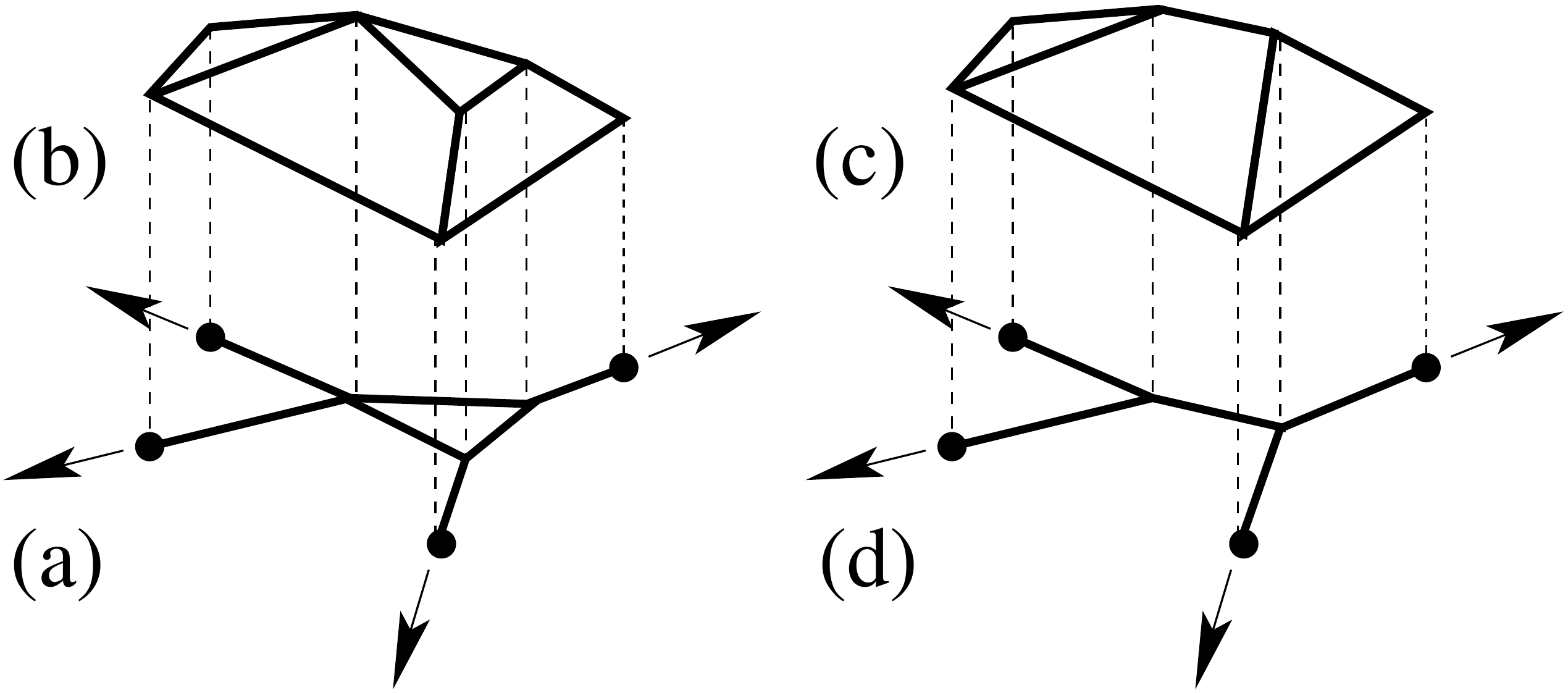}
	\caption{Here in (a) we show a wire web under tension with 4 external nodes at the vertices of a convex quadrilateral, on which the forces are appled, 
        and with three internal nodes. The associated convex polygonal Airy stress function is shown in (b). The simplified convex polygonal Airy stress function
is shown in (c), and the associated simplified network supporting, under tension, the same set of forces as the original network is shown in (d).  }
	\labfig{0.5}
\end{figure}

Given that the web supports, under tension, the forces $\Bt_i$ at the points $\Bx_i$ it will also
support the forces $\Bt_i$ at the points $\tilde{\Bx}_{i}$,
where
  \beq \tilde{\Bx}_{i}=\Bx_i+\Ge\Bt_i, \eeq{1.1a}
and one extends the web, as shown in \fig{1}(b),
by attaching $n$ short wires of length $\Ge>0$ between $\Bx_i$ and $\tilde{\Bx}_{i}$, $i=1,2,\ldots,n$.
The first step is to determine the Airy stress potential $\phi(\Bx)$ in the polygonal ring bounded on one side
by the polygon joining the points $\Bx_i$, and the polygon joining the points $\tilde{\Bx}_{i}$.
When $\Ge$ is sufficiently small there are no wires inside the quadrilateral with vertices $\tilde{\Bx}_{i-1}$, $\Bx_{i-1}$, $\Bx_i$, $\tilde{\Bx}_{i}$
and since the stress vanishes there, the Airy stress potential $\phi(\Bx)$ inside that quadrilateral must be a linear function:
\beq \phi_i(\Bx)=a_ix_1+b_ix_2+c_i. \eeq{1.3}
Continuity of the Airy stress potential $\phi(\Bx)$ at the point $\Bx_i$ then implies
\beq a_{i+1}x_1^{(i)}+b_{i+1}x_2^{(i)}+c_{i+1}=a_ix_1^{(i)}+b_ix_2^{(i)}+c_i, \eeq{1.4}
where $x_1^{(i)}$ and $x_2^{(i)}$ denote the cartesian components of $\Bx_i=(x_1^{(i)}, x_2^{(i)})$.
More generally the line of points $(x_1,x_2)$ where 
$\phi_{i+1}(x_1,x_2)=\phi_i(x_1,x_2)$, i.e. where
\beq a_{i+1}x_1+b_{i+1}x_2+c_{i+1}=a_ix_1+b_ix_2+c_i \eeq{1.4a}
must be parallel to the force vector $\Bt_i$, i.e.
\beq \Bt_i\cdot\bpm a_{i+1}-a_i \\ b_{i+1}-b_i \epm =0. \eeq{1.5}
Across this line $\Grad\phi$ jumps from
\beq \Grad\phi_i=\bpm a_i \\ b_i \epm\quad {\rm to}~\Grad\phi_{i+1}=\bpm a_{i+1} \\ b_{i+1} \epm,
\eeq{1.6}
and this jump 
\beq \Grad\phi_{i+1}-\Grad\phi_i=\bpm a_{i+1}-a_i \\ b_{i+1}-b_i \epm  \eeq{1.7}
can be identified with $\BR_\perp^T\Bt_i$, implying
\beq \Bt_i=\BR_\perp[\Grad\phi_{i+1}-\Grad\phi_i]=\bpm b_{i+1}-b_i \\ a_{i}-a_{i+1} \epm,
\eeq{1.8}
which is consistent with \eq{1.5}. Since a linear function $a_0x_1+b_0x_2$ can be added to $\phi(\Bx)$ without 
changing the stress field $\BGs(\Bx)$ we can assume without loss of generality that
\beq a_1=b_1=c_1=0. \eeq{1.9}
Then \eq{1.8} can be used to determine the remaining coefficients $a_i$ and $b_i$:
\beq \bpm a_{m+1} \\ b_{m+1} \epm = \BR_\perp^T\sum_{i=1}^m \Bt_i= -\BR_\perp\sum_{i=1}^m \Bt_i, \eeq{1.10}
while \eq{1.4} can be used to determine the remaining coefficients $c_i$:
\beq c_{m+1}=\sum_{i=1}^m\Bx_i\cdot[\BR_\perp\Bt_i]. \eeq{1.11}
Of course, since $a_{n+1}$, $b_{n+1}$ and $c_{n+1}$ can be identified with
$a_1$, $b_1$, and $c_1$ which by \eq{1.9} are zero we necessarily have
\beq \sum_{i=1}^n \Bt_i=0,\quad \sum_{i=1}^n\Bx_i\cdot[\BR_\perp\Bt_i]=0, \eeq{1.11a}
 which are the expected conditions for balance of force and torque in the system. We have now determined
the Airy stress function in the polygonal ring. 

\begin{figure}
	\centering
	\includegraphics[width=0.9\textwidth]{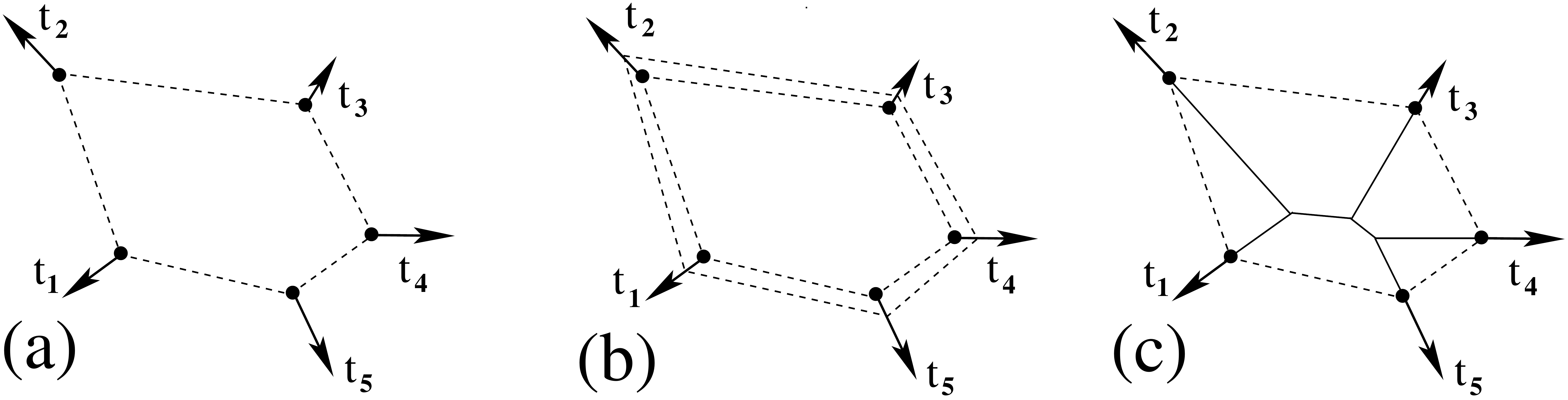}
	\caption{Here (a) shows the forces that are applied at the vertices of the convex polygon; (b) shows the polygonal ring inside which the Airy stress function is first determined; (c) shows the resulting web of wires generated from the lines of discontinuity of slope of the Airy stress function defined by \protect{\eq{1.14}}.}
	\labfig{1}
\end{figure}

Since $\phi(\Bx)$ has non-negative curvature everywhere it must necessarily lie below each tangent plane
$\phi_i(\Bx)$, which provides the necessary constraints that $\phi_i(\Bx_j)\geq \phi_j(\Bx_j)$, or equivalently that
\beq  a_ix_1^{(j)}+b_ix_2^{(j)}+c_i\geq a_jx_1^{(j)}+b_jx_2^{(j)}+c_j, \eeq{1.12}
for all $i$ and $j$. This, when expressed in terms of the forces $\Bt_k$, becomes
\beq \sum_{k=j}^{i-1}(\Bx_k-\Bx_j)\cdot[\BR_\perp\Bt_k]\geq 0.
\eeq{1.13}
where we have assumed $i>j$, if necessary by replacing $i$ by $i+n$ and identifying where necessary $\Bx_k$ and $\Bt_k$ with
$\Bx_{k-n}$ and $\Bt_{k-n}$. The condition \eq{1.13} physically says that the net anticlockwise torque around
the point $\Bx_j$ of the forces $\Bt_k$ summed over any number of consecutive points clockwise past the point
$\Bx_j$ is non-negative. The necessity of condition \eq{0.1} (i.e., \eq{1.13}) in Theorem 1 is thus established.
Similarly, as also implied by \eq{1.13} and \eq{1.11a}, the net clockwise torque around
the point $\Bx_j$ of the forces $\Bt_k$ summed over any number of consecutive points anticlockwise past the point
$\Bx_j$ is non-negative. By taking $i=j+2$, or by taking $i=j+n-2$ and using \eq{1.11a}, we obtain
\beq (\Bx_{j+1}-\Bx_j)\cdot[\BR_\perp\Bt_{j+1}]\geq 0, \quad (\Bx_{j-1}-\Bx_j)\cdot[\BR_\perp\Bt_{j-1}]\leq 0,
\eeq{1.13a}
which is equivalent to the elementary constraint \eq{1.0}.

 In fact this condition is also sufficient: given a set of balanced forces $\Bt_k$, with balanced torques, satisfying
\eq{1.13} then there is a web which supports these forces. The web is easily constructed, as in \fig{0.5}(c), by taking the envelope of the
tangent planes,
\beq \phi(\Bx)=\phi_0(\Bx)\equiv\min_i \phi_i(\Bx)=\min_i\{a_ix_1+b_ix_2+c_i\}, \eeq{1.14}
and placing the web wires where there is a discontinuity in slope in this function.
Clearly this is a function with non-negative curvature and which has the desired tangent planes. This web generated in this way
is an open web with no closed loops, as illustrated in \fig{0.5}(d) and \fig{1}(c). This completes the proof of Theorem 1.

The example of \fig{1}(c) suggests that, among the possible webs that support the given forces, the one given by \eq{1.14}
may be one of the most efficient in the sense of minimizing the total length of all the wire segments. However it is not always
the case as can be seen by the example of forces distributed radially outwards, and all with the same magnitude, around the
boundary on a regular polygon with $m>6$ sides. For example, in \fig{2}(a) we see that for forces at the vertices of a regular 
decagon, the wire geometry generated 
by \eq{1.14} is not as efficient as that in \fig{2}(b), which in turn is not as efficient as that in \fig{2}(c). Other considerations
may be important too in determining the best network to support a given set of forces. For example, following Mitchell \cite{Michell:1904:LEM}, one might
wish to economize the amount of material used while not exceeding some maximum stress threshold. He points out that the Maxwell Lemma
implies that this amount of material is independent of the geometry of the truss network provided one chooses the thicknesses of the struts so
that the cross-sectional stress is constant and at the maximum stress threshold throughout the network. 
As this criterion fails to select the geometry of the network it makes sense to
look for other selection criteria that penalize networks with a complicated geometry. Minimizing the total length of all the wire segments is one such criteria,
another may be minimizing the number of internal nodes, or perhaps a combiniation of these two criteria. Another consideration is that one might
want a network that provides the least obstruction (in some sense that needs to be made precise according to the situation) to the movement of surrounding objects: 
in this respect the network given by \eq{1.14} might be quite good.

\begin{figure}
	\centering
	\includegraphics[width=0.9\textwidth]{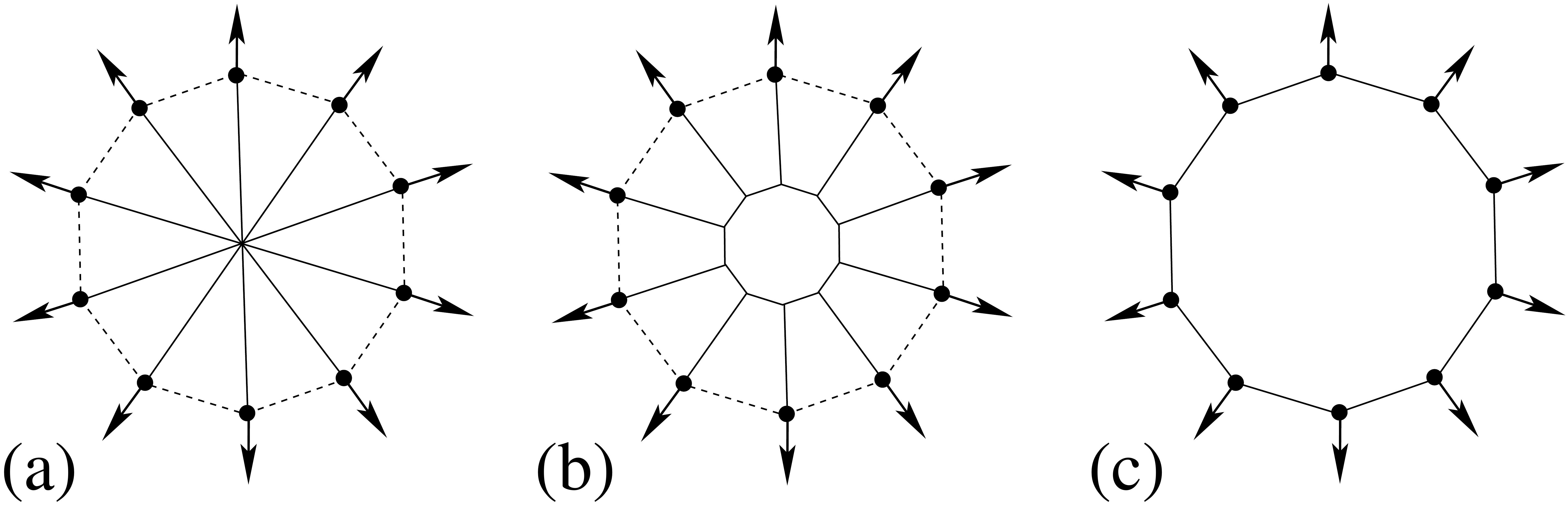}
	\caption{For forces of equal magnitude and pointed radially outwards from the vertices of a regular decagon, the wire geometry generated by \protect{\eq{1.14}}
as shown in (a) is not as efficient, in terms of total length of wire used, as the geometries in (b) and (c)}
	\labfig{2}
\end{figure}

\section{Force cones}
\setcounter{equation}{0}

In this section we consider the set of points $\Bx_1,\Bx_2,\ldots,\Bx_n$ at the vertices of a convex polygon as being fixed, and we 
seek to characterize the complete set of possible loadings that can be supported by a web, not just single loadings.

Consider, for example, the network in \fig{1}(c). It can support the forces $\Bt_1,\Bt_2,\Bt_3,\Bt_4,$ and $\Bt_5$. Since these sum to zero
we can treat $\Bx_5$ as a ground and only keep track of the force quadruplet $\BF=(\Bt_1,\Bt_2,\Bt_3,\Bt_4)$. Schematically, as illustrated
in \fig{3}, we can think of
$\BF$ as a vector in a 8-dimensional space, since each vector $\Bt_i$ has two components. Due to balance of torques, 
\beq \sum_{i=1}^4(\Bx_i-\Bx_5)\cdot[\BR_\perp\Bt_i]=0, \eeq{1.15}
it suffices, in fact, to represent $\BF$ in a 7-dimensional space.
Up to multiplication by a positive constant, this is the only force 
quadruplet $\BF$ that the network can support. The reason for this is easy to see: since each junction inside the web has exactly 
three wires meeting at it, by balance of forces the tension in one wire uniquely determines the tension in the other two wires 
that meet it. Applying this to each junction we see that the tensions in the entire network are uniquely determined once we know
the tension in one wire element, i.e. the tensions in the entire network, and hence the boundary forces $\BF=(\Bt_1,\Bt_2,\Bt_3,\Bt_4)$,
are uniquely determined up to a positive multiplicative factor. 

\begin{figure}
	\centering
	\includegraphics[width=0.9\textwidth]{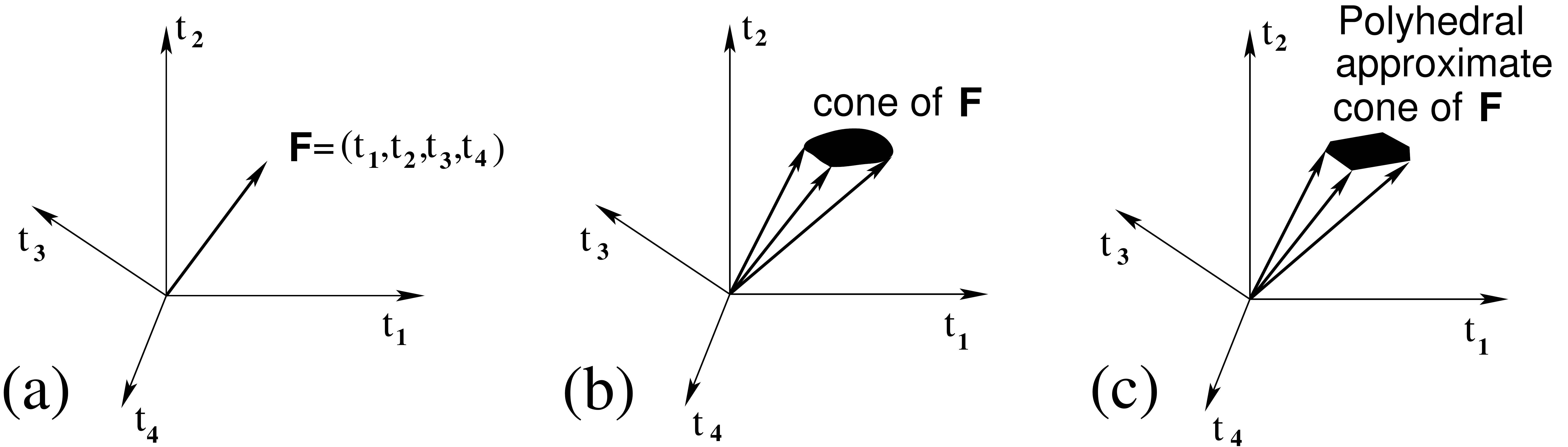}
	\caption{A schematic representation of the force quadruplet $\BF=(\Bt_1,\Bt_2,\Bt_3,\Bt_4)$ in an 8-dimensional space is shown in (a). In general a 
web can support force multiplets $\BF$ that lie in some convex cone, as sketched in (b). To build a web that supports, and only supports, a desired cone of forces
one may first approximate the cone by a polyhedral cone as in (c).}
	\labfig{3}
\end{figure}

Of course there are many networks that support more than one loading. An example is that shown in \fig{2}(a). We are free to choose any forces
$\Bt_1$, $\Bt_2, \ldots,\Bt_{10}$ (numbered clockwise) that point radially outwards from the center of the decagon, provided only that
$\Bt_1+\Bt_6=0$, $\Bt_2+\Bt_7=0$, $\Bt_3+\Bt_8=0$, $\Bt_4+\Bt_9=0$, and $\Bt_5+\Bt_{10}=0$. This degeneracy is removed in \fig{2}(b) and \fig{2}(c) as in those configurations 
each junction inside the web has exactly three wires meeting at it. More generally, as illustrated in \fig{4},
if the wire network generated by the function in \eq{1.14}, has junctions where more than 3 wires meet, we can modify $\phi(\Bx)$ by 
cleaving off the associated parts of $\phi(\Bx)$, resulting in a network where exactly three wires meet at every junction. This ensures
that it supports only the desired force multiplet $\BF$, and positive multiples thereof.

\begin{figure}
	\centering
	\includegraphics[width=0.6\textwidth]{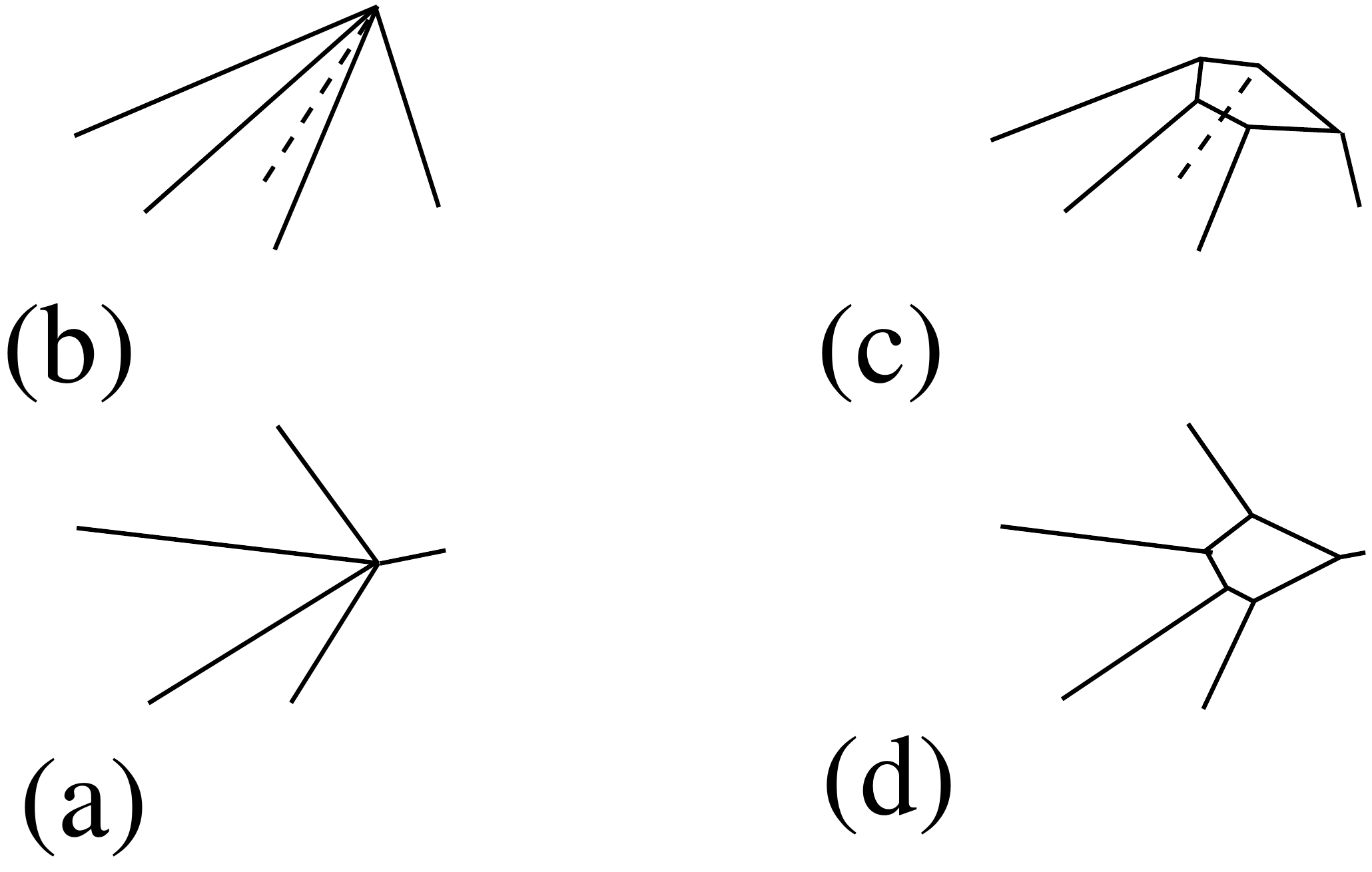}
	\caption{The steps required to remove degeneracy in a web, obtained from \protect{\eq{1.14}} or otherwise,
and thus ensure that it can only support a single force multiplet $\BF$. If the web has any junction where more than three wires meet,
as illustrated in (a), we consider the associated Airy stress function $\phi(\Bx)$ near that point (corresponding to the
desired $\BF$), as in (b), and we cleave it as in (c). This replaces the single junction by a set of junctions, as in (d),
at each of which exactly three wires meet.}
	\labfig{4}
\end{figure}

If a network supports two force multiplets $\BF_1$ and $\BF_2$ then it will also support any weighted average $w\BF_1+(1-w)\BF_2$,
where $w\in [0,1]$, with the corresponding Airy stress function being a weighted average of the two. Thus the set $\CF$ of
possible force multiplets that any given network (with all wires under tension) can support is necessarily a convex cone
as sketched in \fig{3}(b).

Any force multiplet $\BF=(\Bt_1,\Bt_2,\ldots,\Bt_{n-1})\in\CF$ must necessarily satisfy \eq{1.13}, 
with $\Bt_n=-(\Bt_1+\Bt_2+\ldots+\Bt_{n-1})$. Conversely, we may ask the question:
Given any convex cone $\CF$ such that any $\BF=(\Bt_1,\Bt_2,\ldots,\Bt_{n-1})\in\CF$ satisfies \eq{1.13},
can we design a web that supports all those force multiplets in $\CF$ but none other? We can
answer this approximately. Since $\CF$ is convex we can approximate it by a polyhedral cone, as in \fig{3}(c),
that consists of all linear sums with positive coefficients of a finite set of $h$
force multiplets $\BF_1,\BF_2,\ldots,\BF_h$, each satisfying \eq{1.13}. Then we construct
webs that support each force multiplet $\BF_i$, and only that force multiplet and positive
multiples thereof. Finally we superimpose all the $h$ webs. An example of this superposition in the case $h=2$ is given in \fig{5}.

\begin{figure}
	\centering
	\includegraphics[width=0.9\textwidth]{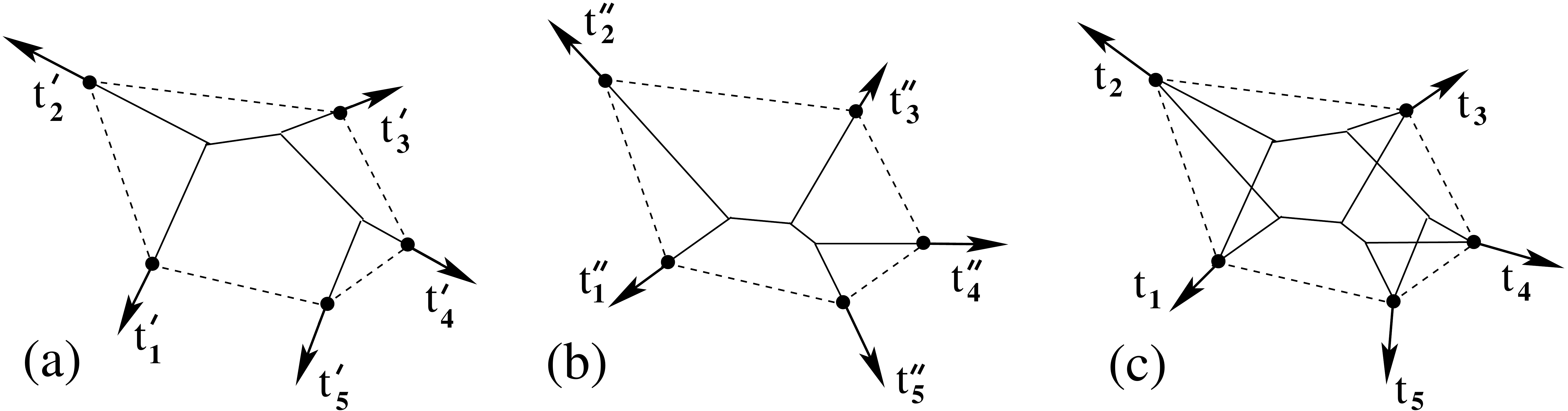}
	\caption{Given a web that supports a force quadruplet $\BF'=(\Bt_1',\Bt_2',\Bt_3',\Bt_4')$ as in (a), and given a web that supports a force quadrulet 
$\BF''=(\Bt_1'',\Bt_2'',\Bt_3'',\Bt_4'')$ as in (b), we can superimpose them, as in (c), to obtain a web that supports force quadruplets $\BF=(\Bt_1,\Bt_2,\Bt_3,\Bt_4)$ 
that are any linear combination $\BF=\Ga'\BF'+\Ga''\BF''$ of $\BF'$ and $\BF''$ with non-negative coefficients $\Ga'$ and $\Ga''$.}
	\labfig{5}
\end{figure}

Of course when we superimpose the networks the wires will cross. If we wish to avoid these crossings, and thus remain in a strictly two-dimensional setting, we
may simply join the wires at the crossing point. Suppose the crossing point is taken as the origin. Then if the wires intersect points $\Ge\Bx_0$, $-\Ge\Bx_0$,
$\Ge\By_0$, $-\Ge\By_0$, with $\By_0$ not parallel to $\Bx_0$, balance of forces at the junction requires the tension in the wire between the origin and $\Ge\Bx_0$ to be the same as the tension in the wire 
between the origin and $-\Ge\Bx_0$. Similarly, balance of forces requires the tension in the wire between the origin and $\Ge\By_0$ to be the same as the tension in the 
wire between the origin and $-\Ge\By_0$. Thus the response with the wires joined at the origin is exactly the same as if they just crossed and were not joined:
the tensions cannot be distributed any other way when only two wires cross at a point. 

It may also be occasionally the case that when we superimpose the networks that a segment of one wire in one network lies exactly on top of another segment of wire
in another network. To avoid that we can choose slightly different force multiplets $\BF_1,\BF_2,\ldots,\BF_h$ that still approximate the desired cone $\CF$. This can
also be done to avoid the case where a junction in one wire network lies exactly on top of the junction in another network-- alternatively, to avoid this,
one may make an operation like that in \fig{4} to replace a junctions in one network by junctions in other places.  
    
This completes the proof of Theorem 2.

\section{Forces at an arbitrary collection of points in the plane}
\setcounter{equation}{0}

Let us first consider the case where forces  $\Bt_2$, $\Bt_3$,..., $\Bt_n$
are applied respectively at $n-1$ points $\Bx_2$, $\Bx_3$,..., $\Bx_n$ that lie
on the vertices of a convex polygon, and an additional force $\Bt_1$ is applied at a point $\By_1$ at a point in the interior
of the polygon. An example is studied in \fig{6}.
If we do have a web that supports 
these forces then we can pick a value of the parameter $s>0$ such that
\beq \Bx_1=\By_1+s\Bt_1, \eeq{2.1}
together with $\Bx_2$, $\Bx_3$,..., $\Bx_n$  lie on the vertices of a convex polygon with $n$ sides. 
We renumber the points, aside from $\Bx_1$, so that their numbering increases consecutively as one goes
clockwise around the boundary of this $n$ sided polygon from $\Bx_1$.
We now join $\Bx_1$ and $\By_1$ 
with a wire and replace the force $\Bt_1$ at $\By_1$ by the same force $\Bt_1$ at $\Bx_1$. Then we can apply the same
analysis as in the preceding section and we deduce that the condition \eq{1.13} necessarily must still hold. 
Notice that the $a_i$ and $b_i$ only depend
on the forces $\Bt_j$, while the $c_i$ only depend on the torques $\Bx_j\cdot[\BR_\perp^T\Bt_j]$ and thus are insensitive to the value
of $s$. Hence in this condition \eq{1.13} we are free to replace $\Bx_1$ by $\By_1$.

Whether this condition is also sufficient is a bit more delicate. We can of course construct the function $\phi(\Bx)$ 
given by \eq{1.14} and the associated open web. The first question is then whether the point $\By_1$ lies on the web wire
of the associated web that directly attaches to the point  $\Bx_1$. If it does then we can just cut the portion of web wire
between $\Bx_1$ and  $\By_1$ and replace the force $\Bt_1$ at $\Bx_1$ by the same force $\Bt_1$ at $\By_1$,
and we are finished. 

If it does not, then we can still obtain necessary and sufficient conditions analogous to Theorem 1, for determining whether or not a given set of forces
can be supported a network having in particular the force $\Bt_1$ at the point $\By_1$. While we currently do not see any easy way to explicitly
write down the needed inequalities, we can determine whether a set of forces can be supported by any network and if so we can construct an appropriate network. 
This can be done by following the numerical algorithm described in the appendix. The necessary and sufficient conditions are that this algorithm yields
the desired network. 

To get some insight into this algorithm, recall that the essential idea
behind Theorem 1, as conveyed in \fig{0.5}, was that the polygonal Airy surface could be replaced by a simpler Airy surface formed by the tangent planes
at the boundary of the convex polygon (at the vertices of which the forces are applied). The essential idea is now similar. If we do have a web that supports
the desired forces, with a wire going from $\Bx_1$ to $\By_1$ then we can replace the associated  polygonal Airy surface by a simpler
polygonal Airy surface formed by the tangent planes at the boundary of the convex polygon intersected with the tangent planes to the ridgeline
associated with the wire going from $\Bx_1$ to $\By_1$. 
As the tension in the wire is constant the angle between these latter tangent planes is fixed by the tension. Thus the tangent planes to the ridgeline associated with
the wire going from
$\Bx_1$ to $\By_1$ can be viewed as a collection of roofs, with a known discontinuity of slope at each roof ridgeline, where each roof (when extended) lies above the value
of the Airy stress function at the vertices $\Bx_1, \Bx_2$,..., $\Bx_n$ of the convex polygon. A change from one ``roof'' to another occurs when 
the wire going from $\Bx_1$ to $\By_1$ crosses another wire in the network. A further simplification of the polygonal Airy surface can be made
by lowering and if need be rotating each roof (while keeping the discontinuity in slope across the ridgeline fixed and keeping the downwards projection 
of the ridgeline coincident with the wire from $\Bx_1$ to $\By_1$) until it touches
the values of the original Airy stress function at two of the vertices $\Bx_1, \Bx_2$,..., $\Bx_n$ of the convex polygon. (The two degrees of freedom
associated with lowering and rotating allow us to match the Airy stress function at two vertices). This leads to the algorithm reported in the appendix.

Now consider the more general case in which forces are applied at $n-h$ points $\Bx_i$ that form the vertices of a convex polygon, and 
forces are applied at an additional $h>1$ points $\By_j$ inside the convex polygon. Now for each point $\By_j$ inside the convex polygon
we choose a constant $s_j>0$ such that the set of $h$ points
\beq \Bx_j=\By_j+s_j\Bt_j, \eeq{2.6}
together with the original $n-h$ points $\Bx_i$ form a convex $n$ sided polygon. We renumber the points so that they increase consecutively as one goes
clockwise around the boundary of this $n$ sided polygon. Then the condition \eq{1.13} must again hold, and must still hold if we replace
the $\Bx_j$ with $\By_j$ for all points $\By_j$ inside the original $(n-h)$-sided polygon. The easiest case is if the web associated with
\eq{1.14} has $\By_j$ on the wire that connects with $\Bx_j$, for all the $h$ interior points $\By_j$. Then we just cut each section
of wire between $\Bx_j$ and $\By_j$ and replace the force $\Bt_j$ at $\Bx_j$ by the same force $\Bt_j$ at $\By_j$. 
If the easiest case does not apply then some suggestions for how to proceed are made in the appendix. Aside from the easiest case, the situation is quite complex 
and probably best tackled numerically.

\section{Truss structures with elements under compression or tension that can support an arbitrary collection of forces at the boundary of a convex polygon.}
\setcounter{equation}{0}

Suppose we wish to construct a truss structure that supports forces $\Bt_i$ at points  $\tilde{\Bx}_{i}$ that are the vertices of a convex polygon
and which only supports this loading. The only constraint on the forces
are that they are balanced and have zero net torque, i.e., that they satisfy \eq{1.11a}. Explicit routes to constructing such a structure,
even for arbitrarily placed points $\tilde{\Bx}_{i}$, has been given in three dimensions by Camar Eddine and Seppecher \cite{Camar:2003:DCS},
and in two-dimensions by Guevara-Vasquez, Milton, and Onofrei \cite{Vasquez:2011:CCS}. Their constructions proceed by induction, and are
quite complicated. Here, in the
two-dimensional case, we show that when the points  $\tilde{\Bx}_{i}$ are the vertices of a convex polygon, the same result can 
be obtained very easily, by essentially just superimposing two truss networks: one with all its elements under compression and the second with all  its
elements under tension.

Clearly the construction is possible
if for some small value of a positive parameter $\Ge$ we can find a truss structure that supports the forces $\Bt_i$ at the points $\Bx_i$,
where $\Bx_i$ is given by \eq{1.1a}. We take a point $\Bx_0$ which is a weighted average of the points $\Bx_i$,
\beq \Bx_0=\sum_{i=1}^nw_i\Bx_i, \eeq{3.1}
where the weights $w_i$ are positive and sum to one. Then, as shown in the example of \fig{7}(b),
we construct a truss structure consisting of $n$ trusses connecting the point $\Bx_0$ 
to the points $\Bx_i$, $i=1,2,\ldots,n$. This truss structure clearly supports forces $\Bt'_i=\Gl\Bs_i$ where
\beq \Bs_i=w_i(\Bx_0-\Bx_i), \eeq{3.2}
and the trusses will all be under compression if $\Gl$ is positive. The weights $w_i>0$ are required to be chosen so $\Bs_j$ and $\Bt_j$
are not parallel for any $j=1,2,\ldots,n$. Having obtained this structure we now ensure that it only supports the force
multiplet $\BF'=(\Bt'_1,\Bt'_2,\ldots,\Bt'_{n-1})$, and positive multiples thereof,
by replacing the spoke structure issuing from $\Bx_0$ by a ring structure, as in \fig{7}(c). 

We then construct, as in \fig{7}(d), a web of wires, or a truss structure, that supports and only supports,
the forces $\Bt''_i=\Bt_i-\Bt'_i$ at the points $\Bx_i$ and is such that all the trusses in this structure
(or wires in the web) are under tension. This requires that $\Gl$ be chosen sufficiently large that \eq{1.13} holds when the $\Bt_i$ are replaced
by the forces $\Bt''_i$:
\beq \sum_{k=j}^{i-1}(\Bx_k-\Bx_j)\cdot[\BR_\perp{\Bt''}^{(k)}]\geq 0,
\eeq{3.3}
or equivalently 
\beq \sum_{k=j}^{i-1}(\Bx_k-\Bx_j)\cdot[\BR_\perp\Bt_k]\geq \Gl \sum_{k=j}^{i-1}(\Bx_k-\Bx_j)\cdot[\BR_\perp\Bs_k],
\eeq{3.4}
where the quantity on the right is negative when $\Gl>0$ and $j+n>i>j+1$
because the truss structure that supports the forces $\Bt'_i=\Gl\Bs_i$ at the
points $\Bx_i$ has all its truss elements under compression, and thus must satisfy the reverse inequality to \eq{1.13} when the $\Bt_i$ are replaced
by the forces $\Bt'_i$. The parameter $\Gl>0$ must be chosen so this inequality holds for all
$i$ and $j$ with $i>j$, i.e.,
\beq \Gl>\max_{\tiny{\begin{matrix} i,j\\j+n>i>j+1 \end{matrix}}} \left\{\frac{\sum_{k=j}^{i-1}(\Bx_k-\Bx_j)\cdot[\BR_\perp\Bt_k]}
{\sum_{k=j}^{i-1}(\Bx_k-\Bx_j)\cdot[\BR_\perp\Bs_k]}\right\},
\eeq{3.5}
where we have excluded the case in which $i=j+1$, since then \eq{3.4} is trivially satisfied since both sides are zero.

Finally, short truss segments are joined from the points $\Bx_i$ to the points $\tilde{\Bx}_{i}$, and the force $\Bt_i$ at
$\Bx_i$ is replaced by the same force $\Bt_i$ at $\tilde{\Bx}_{i}$. The addition of these small truss segments ensures that the
structure only supports one loading. 

\begin{figure}
	\centering
	\includegraphics[width=0.9\textwidth]{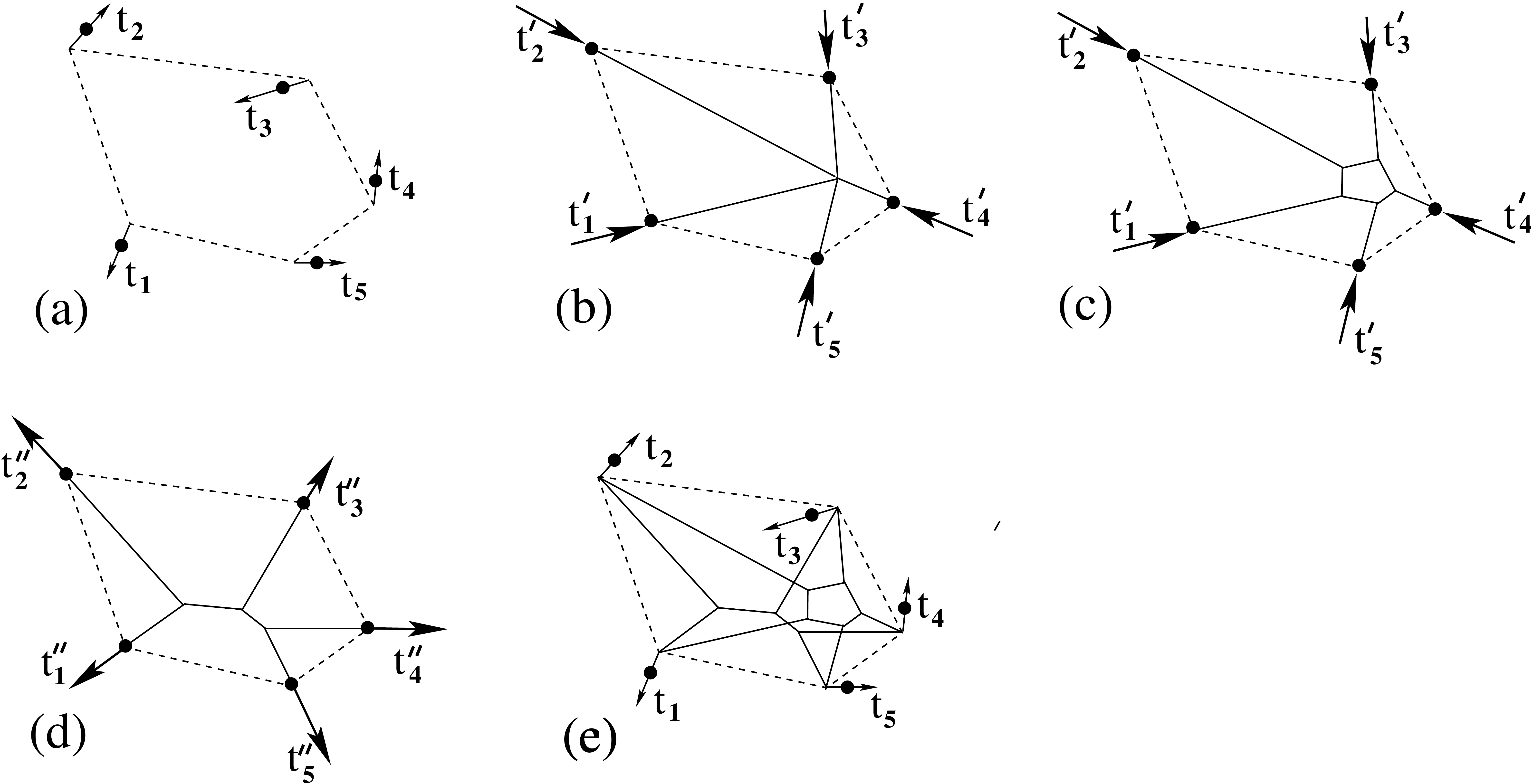}
	\caption{Schematic illustration of the procedure for producing a truss structure that supports at the $n$ vertices of a strictly convex polygon 
one and only force multiplet $\BF=(\Bt_1,\Bt_2,\ldots,\Bt_{n-1})$ that, aside from the constraint of balanced torque, is completely arbitrary.
The first step is to move the points $\tilde{\Bx}_i$ where we apply the forces by a small distance $|\Ge\Bt_i|$ backwards opposite to the direction of these forces.
If $\Ge$ is small enough these new points $\Bx_i$ are still the vertices of a convex polygon, denoted by the dashed outline in (a). The next step
is to take a central junction that is weighted average of the new polygon vertices, and construct a radial truss under compression, 
that supports appropriately large forces $\Bt_1',\Bt_2',\ldots,\Bt_n'$ where $\Bt'_i=\Gl\Bs_i$, $\Gl$ is a positive scaling factor, and $\Bs_i$
is given by the formula \protect{\eq{3.2}}. We then modify this truss structure as in (c) so that it only supports this force multiplet, and positive
multiples thereof.
We then look for a truss structure with all its elements under tension that supports the forces $\Bt_1'',\Bt_2'',\ldots,\Bt_n''$ where
$\Bt''_i=\Bt_i-\Bt'_i$ for all $i$. This is always possible if $\Gl$ is sufficiently large and we can use the Airy function \protect{\eq{1.14}} to
generate the truss structure, as in (d), which may need to be modified in the way indicated in \protect{\fig{4}} to ensure it only supports this
force multiplet, and positive
multiples thereof. Then we superimpose the two truss structures as in (e), with the desired net forces $\Bt_i=\Bt'+\Bt''$ applied now at the
desired points $\tilde{\Bx}_i$, rather than at the vertices $\Bx_i$ of the new polygon.}
\labfig{7}
\end{figure}

\section{Three-dimensional webs}
\setcounter{equation}{0}

Three-dimensional webs are of course much harder to analyse. Some constraints on the forces a web can support can be obtained by projecting
the three-dimensional web onto a two-dimensional space. So, for example, we can consider the two-dimensional web obtained by ``photographing''
the three-dimensional web from above, as illustrated in \fig{8}. If the web is such that it supports forces 
\beq \BT_i=\bpm t^{(i)}_1 \\ t^{(i)}_2 \\ t^{(i)}_3 \epm \quad {\rm applied~at~points}\quad \BX_i=\bpm x^{(i)}_1 \\ x^{(i)}_2 \\ x^{(i)}_3 \epm,
\eeq{4.1}
then the projected two-dimensional web will support forces
\beq  \Bt_i=\bpm t^{(i)}_1 \\ t^{(i)}_2 \epm \quad {\rm applied~at~points}\quad \Bx_i=\bpm x^{(i)}_1 \\ x^{(i)}_2 \epm.
\eeq{4.2}
Thus, for example, at a junction of $p$-wires in the three-dimensional web, if we have balance of the forces $\BF^{(\Ga)}$ that each wire
exerts on the junction,
\beq \sum_{\Ga=1}^p\BF^{(\Ga)}=0,\quad {\rm where}~\BF^{(\Ga)}=\bpm f^{(\Ga)}_1 \\ f^{(\Ga)}_2 \\ f^{(\Ga)}_3 \epm,
\eeq{4.3}
then we will have balance of the forces $\Bf^{(\Ga)}$ that each projected wire exerts on the junction of the projected two-dimensional web:
\beq  \sum_{\Ga=1}^p\Bf^{(\Ga)}=0,\quad {\rm where}~\Bf^{(\Ga)}=\bpm f^{(\Ga)}_1 \\ f^{(\Ga)}_2 \epm.
\eeq{4.4}
Thus the three-dimensional forces $\BT_i$ must be such that the condition \eq{1.13} holds, once we have appropriately renumbered the 
points, and the corresponding forces. Recall that to do this renumbering we need to take the convex hull of the points 
$\Bx_i$ and for points in the interior replace them by points $\Bx_i+s_i\Bt_i$, where $s_i>0$, that are just outside the original 
convex hull. Then one numbers the points going clockwise around the boundary of the new convex hull. 

\begin{figure}
	\centering
	\includegraphics[width=0.3\textwidth]{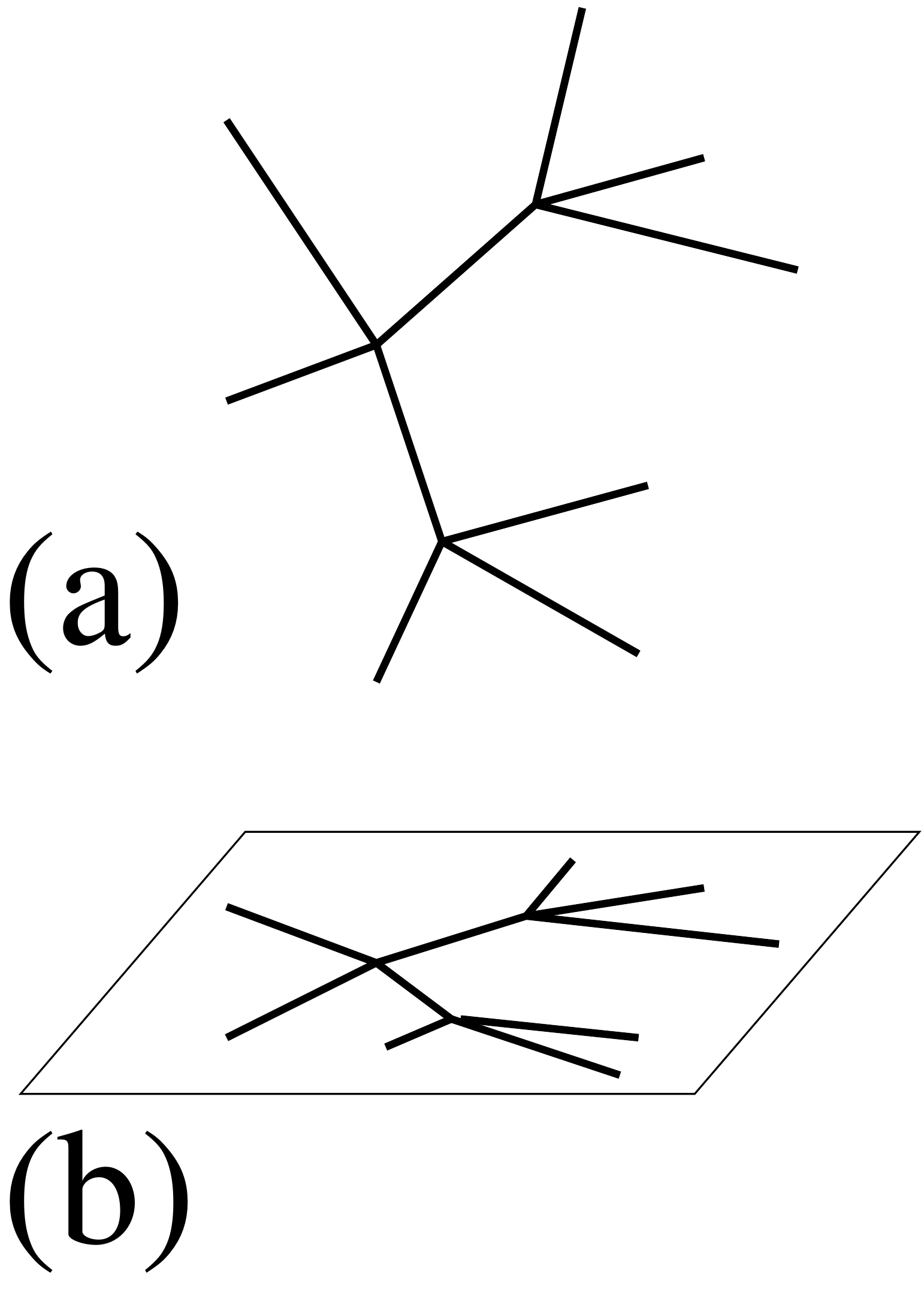}
	\caption{Schematic depiction of how a three-dimensional web under tension, such as (a), has associated two-dimensional webs under tension, such as (b),
that can be obtained by projection onto a plane. In this case
the projection is downwards onto a horizontal plane. In this example, the three-dimensional web can only support one set of applied forces, while the two-dimensional web can support
more than one set of applied forces.}
	\labfig{8}
\end{figure}

This renumbering makes the explicit constraint on the $\BT_i$ not so easy to write down. A simpler, but perhaps weaker, condition can be obtained by
attaching wires to all points $\BX_i$ where we apply forces that go infinity in the direction of $\BT_i$. Then it is the values
of $\Bt_i/|\Bt_i|$ going clockwise around the unit circle that determine the ordering. Of course this ordering will depend
on which projection we choose to take.

There are other, possibly additional, constraints on three-dimensional web. For example, suppose there is a strain field
\beq \BGe_0(\Bx)=[\Grad\Bu_0(\Bx)+(\Grad\Bu_0(\Bx))^T]/2, \eeq{4.5}
such that $\BGe_0(\Bx)$ is defined and positive semidefinite, say within the convex hull $\CC$ of the points 
$\BX_i$. For example one could take $\Bu_0=\Grad\psi(\Bx)$ and require that $\psi(\Bx)$ has positive semidefinite
curvature $\Grad\Grad\psi(\Bx)=\BGe_0(\Bx)$. Then we have the inequality
 \beq 0\leq \int_{\CC}\Tr[\BGe_0(\Bx)\BGs(\Bx)]\,d\Bx=\sum_i\BT_i\cdot\Bu_0(\BX_i).
\eeq{4.6}
Of course there are similar inequalities in the two-dimensional case, but when the points $\Bx_i$ are at the vertices of a 
convex polygon the analysis in Section 2, shows that they
do not provide any restrictions that are additional to the constraints \eq{1.13} as these are in fact necessary and sufficient.
Thus, in the three-dimensional case, it is unclear whether \eq{4.6} provides additional restrictions or not.

We observe that a three-dimensional web geometry such as that in \fig{8}(a) has an important feature. Each junction inside the web has exactly 
four wires meeting at it, and so by balance of forces the tension in one wire uniquely determines the tension in the other three wires 
that meet it. Applying this to each junction we see that the tensions in the entire network are uniquely determined once we know
the tension in one wire element, i.e. the tensions in the entire network, and hence the boundary forces $(\BT_1,\BT_2,\ldots, \BT_8)$,
are uniquely determined up to a positive multiplicative factor. Thus this web can only support
one force multiplet, up to a positive multiplicative factor. This same feature accounts for why pentamode materials,
introduced in \cite{Milton:1995:WET,Sigmund:1995:TMP} and experimentally realized in \cite{Kadic:2012:PPM} essentially support only one stress field.

\section*{Acknowledgements}
G.W. Milton thanks the National Science Foundation for support through grant DMS-1211359. Lorenzo Bardella and Ole Sigmund are thanked for providing some useful references,
Ornella Mattei is thanked for spotting some corrections that needed to be made, and Pierre Seppecher is thanked for stimulating discussions.
The two anomomous referees are thanked for their helpful comments.

\section{Appendix: Constructing suitable web networks when the forces are applied at points which are not the vertices of a convex polygon}
\setcounter{equation}{0}

Let us first consider the case where forces  $\Bt_2$, $\Bt_3$,..., $\Bt_n$
are applied respectively at $n-1$ points $\Bx_2$, $\Bx_3$,..., $\Bx_n$ that lie
on the vertices of a convex polygon, and an additional force $\Bt_1$ is applied at a point $\By_1$ at a point in the interior
of the polygon. The point $\Bx_1$ given by \eq{2.1} is such that all $n$ points $\Bx_1$, $\Bx_2$,..., $\Bx_n$ are the vertices (renumbered
clockwise) of a new convex polygon. We now provide an algorithm for constructing a desired web supporting the force $\Bt_1$ at $\By_1$
and  the forces $\Bt_2$, $\Bt_3$,..., $\Bt_n$ at the $n-1$ points $\Bx_2$, $\Bx_3$,..., $\Bx_n$, if such a web exists. The first step
is to check if the web associated with $\phi_0(\Bx)$ given by \eq{1.14} has $\By_j$ on the wire that connects with $\Bx_j$. If it
does, then we are finished as we can just cut the portion of web wire
between $\Bx_1$ and  $\By_1$ and replace the force $\Bt_1$ at $\Bx_1$ by the same force $\Bt_1$ at $\By_1$.
If it does not, as in the example of \fig{6}(c), then we have to move on to the next step.

The next step is to extend the ridgeline of the function $\phi_0(\Bx)$ going from  the point $(\Bx_1, \phi_0(\Bx_1))$. We want the ridgeline to remain
vertically below the linear extension of the ridgeline that goes from $(\Bx_1, \phi_0(\Bx_1)$, and we want the jump in $\Grad\phi$ across the ridgeline
to remain the same, as it corresponds to the stress in the wire. Accordingly let us define the roof function
\beqa r_1(\Bx,\Ga_1,\Gb_1,\Gg_1) & = &  \Ga_1x_1+\Gb_1 x_2+\Gg_1+ \min\{\phi_2(\Bx), \phi_1(\Bx)\}\nonum
                                & = &\Ga_1x_1+\Gb_1 x_2+\Gg_1+\min\{a_2x_1+b_2x_2+c_2, a_1x_1+b_1x_2+c_1\}.\nonum
&~&
\eeqa{2.2}
By appropriately adjusting $\Ga_1$, $\Gb_1$, and $\Gg_1$ there are three operations we can do on the roof: 
\begin{enumerate}
\item Raising or lowering the roof by increasing or decreasing $\Gg_1$;
\item Swaying the roof by keeping the ridgetop line fixed and rotating the roof about this line (but not allowing either roof faces to move 
beyond a vertical orientation);
\item Tilting the roof by increasing the angle $\Gt$ between the rooftop ridgeline and the rooftop ridgeline of $r_0(\Bx,0,0,0)$
\end{enumerate}
For a given tilt $\Gt$ we can lower the roof and sway it until it just touches at least two of the base points $(\Bx_j, \phi_0(\Bx_j))$
$j=2,3,\ldots,n$, but remains above the other base points, and cannot be lowered further or swayed. This then defines the roof function 
$r_1(\Bx,\theta)$. Now we consider
\beq \phi(\Bx)=\min\{\phi_0(\Bx), r_1(\Bx,\theta)\}, \eeq{2.3}
where $\phi_0(\Bx)$ is defined by \eq{1.14}. We increase the tilt $\theta$ until the projection of the ridgeline down on the $\Bx$-plane
passes through the desired point $\By_1$. If this is impossible, then there is no web that supports the forces. If it is possible, and no
breaks have occurred in the ridgeline, then in the associated web we cut the portion of web wire
between $\Bx_1$ and  $\By_1$ and replace the force $\Bt_1$ at $\Bx_1$ by the same force $\Bt_1$ at $\By_1$,
and we are finished. Then the web has one closed loop, as in the example of \fig{6}(c),
and the force $\Bt_1$ acting at $\By_1$ is attached by wire to one of the vertices of this closed loop. It is 
conceivable that a break has occurred in the ridge line, and in that case to close the break we may need to take
\beq \phi(\Bx)=\min\{\phi_0(\Bx), r_1(\Bx,\theta_1), r_1(\Bx,\theta_2),\ldots, r_1(\Bx,\theta_q)\},
\eeq{2.4}
with appropriate values of $\theta_1<\theta_2<\ldots<\theta_q$, that are clearly non-unique, and chosen to close the break, or breaks, in the ridgeline.
If this is impossible, then there is no web that supports the forces. 

\begin{figure}
	\centering
	\includegraphics[width=0.9\textwidth]{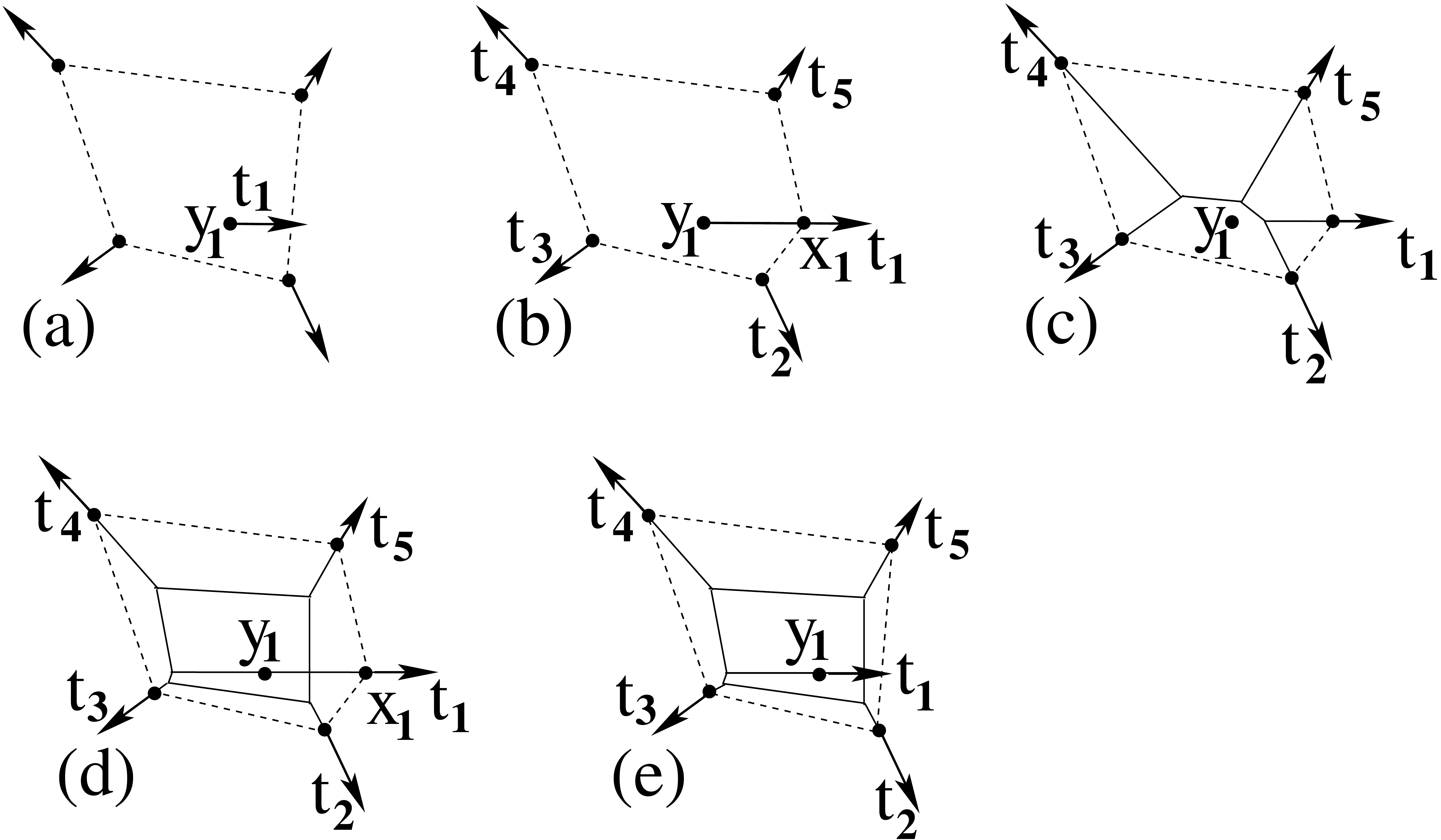}
	\caption{Schematic illustration of the basic procedure when a force $\Bt_1$ is applied at a point $\By_1$ inside the convex hull of the points
at which the other forces are applied, as shown in (a). The first step, as in (b), is to replace the force at  $\Bt_1$ at $\By_1$ by the force $\Bt_1$ at $\Bx_1$, where
$\Bx_1-\By_1$ is in the direction of $\Bt_1$ and $\Bx_1$ lies just outside the original convex hull. Then the other points and forces are numbered clockwise
around the boundary, beginning with $\Bt_1$ at $\Bx_1$. Then the equality \protect{\eq{1.13}} must hold, and the Airy stress function \protect{\eq{1.14}}
gives a web, shown in (c) that supports these forces. It does not necessarily support the force $\Bt_1$ at $\By_1$ unless the point $\By_1$ is on the wire in this web that goes to the 
point $\Bx_1$. Figure (c) shows the case where $\By_1$ is not on the wire, and accordingly one introduces the roof function $r_1(\Bx,\theta)$, and considers the new Airy stress function
defined by \protect{\eq{2.3}}. This generates the web shown in (d) that also supports the same set of forces, including $\Bt_1$ at $\Bx_1$. Now however $\By_1$ lies on
a wire that goes straight to $\Bx_1$, and which, by our choice of roof function, has the same tension everywhere along the wire. The last step, as in (e) is to cut
the wire between $\By_1$ and $\Bx_1$ and replace the force $\Bt_1$ at $\Bx_1$ with the force  $\Bt_1$ at $\By_1$.}
	\labfig{6}
\end{figure}

Now let us briefly shed some light (but not completely solve) the difficult case when forces are applied at $n-h$ points $\Bx_i$ that form the vertices of a convex polygon, and 
forces are applied at an additional $h>1$ points $\By_j$ inside the convex polygon. From the points $\By_j$ we construct associated points $\Bx_j$ given by \eq{2.6}, that are
chosen so that they, together with the original $n-h$ points $\Bx_i$, are the vertices of a new convex polygon, and renumbered so the numbering goes clockwise around the new 
convex polygon. We assume that the easiest case, where the web associated with $\phi_0(\Bx)$ given by
\eq{1.14} has $\By_j$ on the wire that connects with $\Bx_j$, for all the $h$ interior points $\By_j$, does not hold.
Then it makes sense to introduce the roof functions,
\beqa &~& r_j(\Bx,\Ga_j,\Gb_j,\Gg_j)  =  \Ga_jx_1+\Gb_j x_2+\Gg_j+ \min\{\phi_{j+1}(\Bx), \phi_{j}(\Bx)\} \nonum
 &~&\quad =  \Ga_jx_1+\Gb_j x_2+\Gg_j+\min\{a_{j+1}x_1+b_{j+1}x_2+c_{j+1}, a_jx_1+b_jx_2+c_j\}. \nonum 
&~&
\eeqa{2.7}
A suitable web, if it exists, should be obtained by taking a minimum over a set of functions consisting of $\phi_0(\Bx)$ and an 
appropriate choice of roof functions $r_j(\Bx,\Ga_j,\Gb_j,\Gg_j)$. It is no longer that case that one should lower the roof 
functions as much as possible, as such a lowering could significantly shorten some other ridgelines. 
It seems likely that there could be competition between lengthening one ridge line, and lengthening another ridge line. It may happen for some $j$ and $k$, with $j\ne k$ that the wire between $\Bx_j$ and $\By_j$ crosses
the wire between $\Bx_k$ and $\By_k$ at some point $\Bx_{j,k}$. In that case one should introduce the pyramid roof functions
\beq p(\Bx,\Ga,\Gb,\Gg)=\Ga x_1+\Gb x_2+\Gg+ \min\{r_j(\Bx,0,0,0), r_j(\Bx,\Ga_j,\Gb_j,\Gg_j)\}, \eeq{2.8}
where the three parameters $\Ga_j$, $\Gb_j$, and $\Gg_j$ are chosen so the line of discontinuity in slope of $p(\Bx,\Ga,\Gb,\Gg)$
that meets the ridgeline of $r_j(\Bx,0,0,0)$ and the ridgeline of $r_j(\Bx,\Ga_j,\Gb_j,\Gg_j)$ at the top of the pyramid
corresponds to the wire between $\Bx_k$ and $\By_k$, and the jump in slope across this line corresponds to the 
tension in that wire. Then one should take $\phi(\Bx)$ to be the minimum of $\phi_0(\Bx)$, $p(\Bx,\Ga,\Gb,\Gg)$ and
possibly other appropriate roof functions, where $\Ga$, $\Gb$, and $\Gg$ need to be suitably chosen.
In this scenario one may not need to explicitly use the roof functions $r_k(\Bx,\Ga_k,\Gb_k,\Gg_k)$. In the event that $\Bx_{j,k}$
lies on the ridgeline of $\phi_0(\Bx)$ that reaches $\Bx_j$ it may suffice to take $\Ga=\Gb=\Gg=0$.
\end{document}